\documentclass[prd,preprintnumbers,superscriptaddress,nofootinbib]{revtex4}
\usepackage[T1]{fontenc}
\usepackage{amsmath,amssymb}
\usepackage{epsfig}
\usepackage{graphicx}
\usepackage[usenames,dvipsnames]{color}
\usepackage{slashed}
\usepackage[colorlinks,citecolor=blue]{hyperref}
\usepackage{pdfpages}
\usepackage{epstopdf}
\usepackage{color}
\begin{document}
%\preprint{IP/BBSR/2015-4}
\title{Lepton Portal Limit of Inert Higgs Doublet Dark Matter \\ with Radiative Neutrino Mass}
\author{Debasish Borah}
\email{dborah@iitg.ernet.in}
\affiliation{Department of Physics, Indian Institute of Technology Guwahati, Assam 781039, India}
\author{Soumya Sadhukhan}
\email{soumyas@prl.res.in}
\affiliation{Physical Research Laboratory, Ahmedabad 380009, India}

\author{Shibananda Sahoo}
\email{shibananda@iitg.ernet.in}
\affiliation{Department of Physics, Indian Institute of Technology Guwahati, Assam 781039, India}

\begin{abstract}
We study an extension of the Inert Higgs Doublet Model (IHDM) by three copies of right handed neutrinos and heavy charged leptons such that both the inert Higgs doublet and the heavy fermions are odd under the $Z_2$ symmetry of the model. The neutrino masses are generated at one loop in the scotogenic fashion. Assuming the neutral scalar of the inert Higgs to be the dark matter candidate, we particularly look into the region of parameter space where dark matter relic abundance is primarily governed by the inert Higgs coupling with the leptons. This corresponds to tiny Higgs portal coupling of dark matter as well as large mass splitting within different components of the inert Higgs doublet suppressing the coannihilations. Such lepton portal couplings can still produce the correct relic abundance even if the Higgs portal couplings are arbitrarily small. Such tiny Higgs portal couplings may be responsible for suppressed dark matter nucleon cross section as well as tiny invisible branching ratio of the standard model Higgs, to be probed at ongoing and future experiments. We also briefly discuss the collider implications of such a scenario.
\end{abstract}
\pacs{12.60.Fr,12.60.-i,14.60.Pq,14.60.St}
\maketitle

\section{Introduction}
The observational evidence suggesting the presence of dark matter (DM) in the Universe are irrefutable, with the latest data from the Planck experiment \cite{Planck15} indicating that approximately $27\%$ of the present Universe is composed of dark matter. The observed abundance of DM is usually represented in terms of density parameter $\Omega$ as 
\begin{equation}
\Omega_{\text{DM}} h^2 = 0.1187 \pm 0.0017
\label{dm_relic}
\end{equation}
where $h = \text{(Hubble Parameter)}/100$ is a parameter of order unity. In spite of astrophysical and cosmological evidences confirming the presence of DM, the fundamental nature of DM is not yet known. Since none of the particles in the Standard Model (SM) can fulfil the criteria of a DM candidate, several beyond Standard Model (BSM) proposals have been put forward in the last few decades. Among them, the weakly interacting massive particle (WIMP) paradigm is the most popular one. Such WIMP dark matter candidates can interact with the SM particles through weak interactions and hence can be produced at the Large Hadron Collider (LHC) or can scatter off nuclei at dark matter direct detection experiments like the ongoing LUX \cite{LUX16} and PandaX-II experiment \cite{PandaXII}.

Among different BSM proposals to incorporate dark matter, the inert Higgs doublet model (IHDM) \cite{Barbieri:2006dq,Cirelli:2005uq,LopezHonorez:2006gr} is one of the simplest extensions of the SM with an additional scalar field transforming as doublet under $SU(2)$ and having hypercharge $Y=1$, odd under an imposed $Z_2$ discrete symmetry. As shown by the earlier works on IHDM, there are typically two mass ranges of DM mass satisfying the correct relic abundance criteria: one below the $W$ boson mass and the other around 550 GeV or above. Among these, the low mass regime is particularly interesting due to stronger direct detection bounds. For example, the latest data from the LUX experiment rules out DM-nucleon spin independent cross section above around $2.2 \times 10^{-46} \; \text{cm}^2$ for DM mass of around 50 GeV \cite{LUX16}. In this mass range, as we discuss in details below, the tree level DM-SM interaction through the SM Higgs $(h)$ portal is interesting as it can simultaneously control the relic abundance as well as the DM-nucleon scattering cross section. In this mass range, only a narrow region near the resonance $m_{\text{DM}} \approx m_h/2$ is currently allowed by the LUX data. Though future DM direct detection experiments will be able to probe this region further, it could also be true that the DM-Higgs interaction is indeed too tiny to be observed at experiments. Such a tiny Higgs portal interaction will also be insufficient to produce the correct relic abundance of DM in this low mass regime. This almost rules out the low mass regime of DM in IHDM  $m_{\text{DM}} \lessapprox 70$ GeV.

Here we consider a simple extension of IHDM by singlet leptons (both neutral and charged) odd under the $Z_2$ symmetry such that the inert scalar dark matter can interact with the SM particles through these singlet leptons. This new interaction through lepton portal can revive the low mass regime of inert scalar DM even if future direct detection experiment rules out the Higgs portal interaction completely. The lepton portal interactions can also remain unconstrained from the limits on DM-nucleon interactions. Such a scenario is particularly interesting if LHC finds some signatures corresponding to the low mass regime of inert scalar DM while the direct detection continues to give null results. The dominant lepton portal interactions can explain correct relic abundance, null results at direct detection experiments and also give rise to interesting signatures at colliders. The neutral leptons added to IHDM can also give rise to tiny neutrino masses at one-loop level through scotogenic fashion \cite{ma06}. We discuss the constraints on the model parameters from neutrino mass, DM constraints and also make some estimates of some interesting collider signatures while comparing them with the pure IHDM.

This article is organised as follows. In section \ref{sec1}, we discuss the IHDM and then consider the lepton portal extension of it in section \ref{sec2}. In section \ref{sec3}, we discuss the dark matter related studies followed by our collider estimates in section \ref{sec4}. We finally conclude in section \ref{sec5}.

\section{Inert Higgs Doublet Model}
\label{sec1}
The inert Higgs Doublet Model (IHDM) \cite{Barbieri:2006dq,Cirelli:2005uq,LopezHonorez:2006gr} is an extension of the Standard Model (SM)
by an additional Higgs doublet $\Phi_2$ and a discrete $Z_2$ symmetry under which all SM fields are even while $\Phi_2 \rightarrow -\Phi_2$. This $Z_2$ symmetry not only prevents the coupling of SM fermions to $\Phi_2$ at renormalisable level but also forbids those terms in the scalar potential which are linear or trilinear in $\Phi_2$. Therefore, the second Higgs doublet $\Phi_2$ can interact with the SM particles only through its couplings to the SM Higgs doublet and the electroweak gauge bosons. The $Z$ symmetry also prevents the lightest component of $\Phi_2$ from decaying, making it stable on cosmological scale. If one of the neutral components of $\Phi_2$ happen to be the lightest $Z_2$ odd particle, then it can be a potential dark matter candidate. The scalar potential of the model involving the SM Higgs doublet $\Phi_1$ and the inert doublet $\Phi_2$ can be written as
\begin{equation}
\begin{aligned}
V(\Phi_1,\Phi_2)=  \mu_1^2|\Phi_1|^2 +\mu_2^2|\Phi_2|^2+\frac{\lambda_1}{2}|\Phi_1|^4+\frac{\lambda_2}{2}|\Phi_2|^4+\lambda_3|\Phi_1|^2|\Phi_2|^2\\+\lambda_4|\Phi_1^\dag \Phi_2|^2 + \{\frac{\lambda_5}{2}(\Phi_1^\dag \Phi_2)^2 + \text{h.c.}\},
\end{aligned}
\label {c}
\end{equation}
To ensure that none of the neutral components of the inert Higgs doublet acquire a non-zero vacuum expectation value (vev), $\mu_2^2 >0$ is assumed. This also prevents the $Z_2$ symmetry from being spontaneously broken. The electroweak symmetry breaking (EWSB) occurs due to the non-zero vev acquired by the neutral component of $\Phi_1$. After the EWSB these two scalar doublets can be written in the following form in the unitary gauge.
\begin{equation}
\Phi_1=\begin{pmatrix} 0 \\  \frac{ v +h }{\sqrt 2} \end{pmatrix} , \Phi_2=\begin{pmatrix} H^+\\  \frac{H+iA}{\sqrt 2} \end{pmatrix}
\end{equation}
The masses of the physical scalars at tree level can be written as
\begin{eqnarray}
m_h^2 &=& \lambda_1 v^2 ,\nonumber\\
m_{H^+}^2 &=& \mu_2^2 + \frac{1}{2}\lambda_3 v^2 , \nonumber\\
m_{H}^2 &=& \mu_2^2 + \frac{1}{2}(\lambda_3+\lambda_4+\lambda_5)v^2=m^2_{H^\pm}+
\frac{1}{2}\left(\lambda_4+\lambda_5\right)v^2  , \nonumber\\
m_{A}^2 &=& \mu_2^2 + \frac{1}{2}(\lambda_3+\lambda_4-\lambda_5)v^2=m^2_{H^\pm}+
\frac{1}{2}\left(\lambda_4-\lambda_5\right)v^2.
\label{mass_relation}
\end{eqnarray}
Here $m_h$ is the SM like Higgs boson mass, $m_H, m_A$ are the masses of the CP even and CP odd scalars from the inert doublet. Without loss of generality, we consider $\lambda_5 <0, \lambda_4+\lambda_5 <0$ so that the CP even scalar is the lightest $Z_2$ odd particle and hence a stable dark matter candidate.

The new scalar fields discussed above can be constrained from the LEP I precision measurement of the $Z$ boson decay width. In order to forbid the decay channel $Z \rightarrow H A$, one arrives at the constraint $m_H + m_A > m_Z$. In addition to this, the LEP II constraints roughly rule out the triangular
region \cite{Lundstrom:2008ai}
\[
	m_{H} < 80\ {\rm\ GeV},\quad m_{A} < 100{\rm\ GeV},\quad
	m_{A} - m_{H} > 8{\rm\ GeV}
\]
The LEP collider experiment data restrict the charged scalar mass to $m_{H^+} > 70-90$ GeV \cite{lep1}. The Run 1 ATLAS dilepton limit is discussed in the context of IHDM in Ref.\cite{dilepton_IDM} taking into consideration of specific masses of charged Higgs. Another important restriction on $m_{H^+}$ comes from the electroweak precision data (EWPD). Since the contribution of the additional doublet $\Phi_2$ to electroweak S parameter is always small \cite{Barbieri:2006dq}, we only consider the contribution to the electroweak T parameter here. The relevant contribution is given by \cite{Barbieri:2006dq}
\begin{equation}
\Delta T = \frac{1}{16 \pi^2 \alpha v^2} [F(m_{H^+}, m_{A})+F(m_{H^+}, m_{H}) -F(m_{A}, m_{H})]
\end{equation}
where 
\begin{equation}
F(m_1, m_2) = \frac{m^2_1+m^2_2}{2}-\frac{m^2_1m^2_2}{m^2_1-m^2_2} \text{ln} \frac{m^2_1}{m^2_2}
\end{equation}
The EWPD constraint on $\Delta T$ is given as \cite{honorez1}
\begin{equation}
-0.1 < \Delta T + T_h < 0.2
\end{equation}
where $T_h \approx -\frac{3}{8 \pi \cos^2{\theta_W}} \text{ln} \frac{m_h}{m_Z}$ is the SM Higgs contribution to the T parameter \cite{peskin}. 
%\section{Inert Doublet Model+VLF}
%Let us add a Vector-like Lepton doublet which is $Z_2$ odd.
%%%%%%%%%%%%%%%%%%%%%%%%%%%%%%%%%%%%%%%%%%%
\section{Lepton Portal Extensions of IHDM}
\label{sec2}
As discussed in the introduction, considering lepton portal extensions of IHDM is very well motivated, specially from the origin of neutrino mass, dark matter direct detections and other flavour physics observables in the lepton sector. The inert Higgs doublet of the IHDM can couple to the SM leptons, if the model is suitably extended either by $Z_2$ odd neutral Majorana fermions or by charged vector like leptons, none of which introduce any chiral anomalies.
\begin{figure}[htb]
\centering
\includegraphics[scale=0.75]{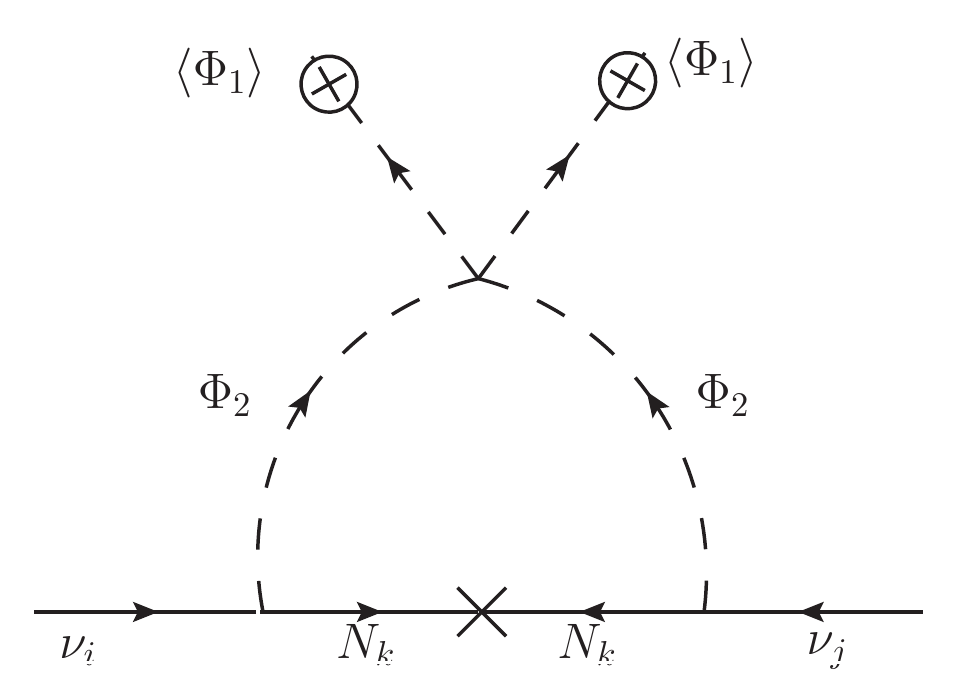}
\caption{One-loop contribution to neutrino mass}
\label{fig1}
\end{figure}
The addition of three copies of neutral heavy singlet fermions $N_i$, odd under the $Z_2$ symmetry leads to the upgradation of the IHDM to the scotogenic model \cite{ma06}. Apart from providing another dark matter candidate in terms of the lightest $N_i$, the model also can explain tiny neutrino masses at one loop level. The relevant interaction terms of these singlet fermions can be written as
\begin{equation}
{\cal L} \supset  M_N N N + \left(Y_{ij} \, \bar{L}_i \tilde{\Phi}_2 N_j  + \text{h.c.} \right) \ . 
\end{equation}

\begin{figure}[htb]
\centering
\includegraphics[scale=0.4]{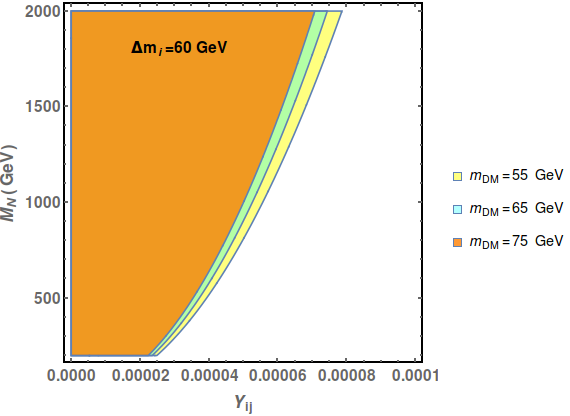}
\caption{Allowed model parameters for neutrino mass generation}
\label{fig1a}
\end{figure}

The Feynman diagram for such one loop neutrino mass is shown in figure \ref{fig1}. Using the expression from \cite{ma06} of one-loop neutrino mass 
\begin{equation}
(m_{\nu})_{ij} = \frac{Y_{ik}Y_{jk} M_{k}}{16 \pi^2} \left ( \frac{m^2_R}{m^2_R-M^2_k} \text{ln} \frac{m^2_R}{M^2_k}-\frac{m^2_I}{m^2_I-M^2_k} \text{ln} \frac{m^2_I}{M^2_k} \right)
\end{equation}
Here $m^2_{R,I}=m^2_{H,A}$ are the masses of scalar and pseudo-scalar part of $\Phi^0_2$ and $M_k$ the mass of singlet fermion $N$ in the internal line. The index $i, j = 1,2,3$ runs over the three fermion generations as well as three copies of $N$. For $m^2_{H}+m^2_{A} \approx M^2_k$, the above expression can be simply written as
\begin{equation}
(m_{\nu})_{ij} \approx \frac{\lambda_5 v^2}{32 \pi^2}\frac{Y_{ik}Y_{jk} }{M_k} =  \frac{m^2_A-m^2_H}{32 \pi^2}\frac{Y_{ik}Y_{jk} }{M_k}
\end{equation}
In this model for the neutrino mass to match with experimentally observed limits ($\sim 0.1$~eV), very tiny Yukawa couplings are required for the right handed
neutrino mass of order of 1 TeV. Taking the mass difference $m_A-m_H=m_{H^{\pm}}-m_H = 60$ GeV, we show the constraints on neutral singlet fermion mass and corresponding Yukawa coupling from correct neutrino mass requirement in figure \ref{fig1a}. It can be seen that for low mass regime of DM, the neutrino mass constraints force the Yukawa couplings to be smaller than $10^{-4}$, too small to have any impact on dark matter relic abundance calculation, to be discussed below. These neutral fermions can also contribute to charged lepton flavour violation (LFV) at one loop involving $N, \Phi^{\pm}_2$. The LFV processes like $\mu \rightarrow e \gamma$ remain suppressed in the SM due to the smallness of neutrino masses.  Such LFV decays like $\mu \rightarrow e \gamma$ are being searched for at experiments like MEG \cite{MEG16}. The latest bound from the MEG collaboration is $\text{BR}(\mu \rightarrow e \gamma) < 4.2 \times 10^{-13}$ at $90\%$ confidence level \cite{MEG16}. However, due to small Yukawa couplings, as required by tiny neutrino mass constraints discussed above, keeps this new contribution to  $\mu \rightarrow e \gamma$ way below this latest experimental bound, as discussed in the recent works \cite{db2, db3}.

Similar to neutral singlet fermions, one can also incorporate charged singlet leptons $\chi_{L,R}$ with hypercharge $Y=2$ and odd under the $Z_2$ symmetry. The relevant  Lagrangian is 
\begin{equation}
{\cal L} \supset  M_{\chi} \bar{\chi}_L \chi_R + Y_{ij} \, \bar{L}_i \Phi_2 \chi_R  + \text{h.c.}
\end{equation}
These leptons can contribute both to dark matter relic abundance as well as LFV decays mentioned above. Since the corresponding Yukawa couplings are not restricted to be small from neutrino mass constraints, they can be sizeable and hence play a non-trivial role in generating DM relic abundance as we discuss below. Such large Yukawa couplings can however give a large contribution to LFV decays like $\mu \rightarrow e \gamma$, with $\chi, \Phi^0_2$ in loop. As shown in a recent work \cite{db4}, the above MEG bound can constrain the product of two relevant Yukawa couplings to be below $10^{-9}$ for $\chi$ mass around 100 GeV-1 TeV, too small to have any impact on DM relic abundance. These strict bounds from MEG can however be evaded by choosing diagonal structure of singlet lepton mass matrix $M_{\chi}$ and relevant Yukawa coupling $Y$. Such a structure can still have non-trivial impact on DM relic abundance, to be discussed below.
%%%%%%%%%%%%%%%%%%%%%%%%%%%%%%%%%%
\section{Dark Matter}
\label{sec3}
The relic abundance of a dark matter particle $\psi$ which was in thermal equilibrium at some earlier epoch can be calculated by solving the Boltzmann equation
\begin{equation}
\frac{dn_{\psi}}{dt}+3Hn_{\psi} = -\langle \sigma v \rangle (n^2_{\psi} -(n^{\text{eqb}}_{\psi})^2)
\end{equation}
where $n_{\psi}$ is the number density of the dark matter particle $\psi$ and $n^{eqb}_{\psi}$ is the number density when $\psi$ was in thermal equilibrium. $H$ is the Hubble expansion rate of the Universe and $ \langle \sigma v \rangle $ is the thermally averaged annihilation cross section of the dark matter particle $\psi$. In terms of partial wave expansion $ \langle \sigma v \rangle = a +b v^2$. Clearly, in the case of thermal equilibrium $n_{\psi}=n^{\text{eqb}}_{\psi}$, the number density is decreasing only by the expansion rate $H$ of the Universe. The approximate analytical solution of the above Boltzmann equation gives \cite{Kolb:1990vq, kolbnturner}
\begin{equation}
\Omega_{\psi} h^2 \approx \frac{1.04 \times 10^9 x_F}{M_{Pl} \sqrt{g_*} (a+3b/x_F)}
\end{equation}
where $x_F = m_{\psi}/T_F$, $T_F$ is the freeze-out temperature, $g_*$ is the number of relativistic degrees of freedom at the time of freeze-out and $M_{Pl} \approx 10^{19}$ GeV is the Planck mass. Here, $x_F$ can be calculated from the iterative relation 
\begin{equation}
x_F = \ln \frac{0.038gM_{\text{Pl}}m_{\psi}<\sigma v>}{g_*^{1/2}x_F^{1/2}}
\label{xf}
\end{equation}
The expression for relic density also has a more simplified form given as \cite{Jungman:1995df}
\begin{equation}
\Omega_{\psi} h^2 \approx \frac{3 \times 10^{-27} \text{cm}^3 \text{s}^{-1}}{\langle \sigma v \rangle}
\label{eq:relic}
\end{equation}
The thermal averaged annihilation cross section $\langle \sigma v \rangle$ is given by \cite{Gondolo:1990dk}
\begin{equation}
\langle \sigma v \rangle = \frac{1}{8m^4_{\psi}T K^2_2(m_{\psi}/T)} \int^{\infty}_{4m^2_{\psi}}\sigma (s-4m^2_{\psi})\surd{s}K_1(\surd{s}/T) ds
\label{eq:sigmav}
\end{equation}
where $K_i$'s are modified Bessel functions of order $i$, $m_{\psi}$ is the mass of Dark Matter particle and $T$ is the temperature.

If we consider the neutral component of the scalar doublet $\Phi_2$ to be the dark matter candidate, the details of relic abundance calculation is similar to the inert doublet model studied extensively in the literature \cite{ma06,Barbieri:2006dq,Majumdar:2006nt,LopezHonorez:2006gr,ictp,borahcline, honorez1,DBAD14}. In the low mass regime $m_H=m_{DM} \leq M_W$, dark matter annihilation into the SM fermions through s-channel Higgs mediation dominates over other channels. As pointed out by \cite{honorez2}, the dark matter annihilations $H H \rightarrow W W^* \rightarrow W f \bar{f^{\prime}}$ can also play a role in the $m_{DM} \leq M_W$ region. Also, depending on the mass differences $m_{H^+}-m_H, m_A-m_H$, the coannihilations of $H, H^+$ and $H, A$ can also play a role in generating the relic abundance of dark matter. The relic abundance calculation incorporating these effects were studied by several groups in \cite{Griest:1990kh, coann_others}. Beyond the W boson mass threshold, the annihilation channel of scalar doublet dark matter into $W^+W^-$ pairs opens up suppressing the relic abundance below what is observed by Planck experiment, unless the dark matter mass is heavier than around 500 GeV, depending on the DM-Higgs coupling. Apart from the usual annihilation channels of inert doublet dark matter, in this model there is another interesting annihilation channel where dark matter annihilates into a pair of neutrinos (charged leptons) through the heavy fermion $N_i$ ($\chi$) in the t-channel. 

Apart from the relic abundance constraints from Planck experiment, there exists strict bounds on the dark matter nucleon cross section from direct detection experiments like Xenon100 \cite{Aprile:2013doa} and more recently LUX \cite{LUX, LUX16}. For scalar dark matter considered in this work, the relevant spin independent scattering cross section mediated by SM Higgs is given as \cite{Barbieri:2006dq}
\begin{equation}
 \sigma_{\text{SI}} = \frac{\lambda^2_L f^2}{4\pi}\frac{\mu^2 m^2_n}{m^4_h m^2_{DM}}
\label{sigma_dd}
\end{equation}
where $\mu = m_n m_{DM}/(m_n+m_{DM})$ is the DM-nucleon reduced mass and $\lambda_L=(\lambda_3+\lambda_4+\lambda_5)$ is the quartic coupling involved in DM-Higgs interaction. A recent estimate of the Higgs-nucleon coupling $f$ gives $f = 0.32$ \cite{Giedt:2009mr} although the full range of allowed values is $f=0.26-0.63$ \cite{mambrini}. The latest LUX bound \cite{LUX16} on $\sigma_{\text{SI}}$ constrains the $\eta_L$-Higgs coupling $\lambda$ significantly, if $\eta_L$ gives rise to most of the dark matter in the Universe. According to this latest bound, at a dark matter mass of 50 GeV, dark matter nucleon scattering cross sections above $1.1 \times 10^{-46} \; \text{cm}^2$ are excluded at $90\%$ confidence level. Similar but slightly weaker bound has been reported by the PandaX-II experiment recently \cite{PandaXII}. We however include only the LUX bound in our analysis. One can also constrain the DM-Higgs coupling $\lambda$ from the latest LHC constraint on the invisible decay width of the SM Higgs boson. This constraint is applicable only for dark matter mass $m_{DM} < m_h/2$. The invisible decay width is given by
\begin{equation}
\Gamma (h \rightarrow \text{Invisible})= {\lambda^2_L v^2\over 64 \pi m_h} 
\sqrt{1-4\,m^2_{DM}/m^2_h}
\end{equation}
The latest ATLAS constraint on invisible Higgs decay is \cite{ATLASinv}
$$\text{BR} (h \rightarrow \text{Invisible}) = \frac{\Gamma (h \rightarrow \text{Invisible})}{\Gamma (h \rightarrow \text{Invisible}) + \Gamma (h \rightarrow \text{SM})} < 22 \%$$
As we will discuss below, this bound is weaker than the LUX 2016 bound.

It should be noted that, there can be sizeable DM-nucleon scattering cross section at one loop level as well, which does not depend on the Higgs portal coupling discussed above. Even in the minimal IHDM such one loop scattering can occur with charged scalar and electroweak gauge bosons in loop \cite{oneloopIDM}. The contributions of such one loop scattering can be kept even below future direct detection experiments like Xenon-1T by choosing large mass differences between the components of the inert scalar doublet \cite{oneloopIDM}. Such large mass splittings also minimise the role of coannihilation between different inert scalar components on the DM relic abundance. This is in the spirit of the present work's motivation, as the DM abundance is primarily determined by the lepton portal couplings, rather than gauge and Higgs portal couplings. Another one loop scattering can occur, in principle, due to the exchange of photons or $Z$ boson. This is possible through an effective coupling of the form $C \partial^{\mu} \Phi^0_2 \partial^{\nu} \Phi^{0\dagger}_2 F_{\mu \nu}$ with $C$ being the loop factor \cite{oneloopLepton}. However, since we have broken the degeneracy of our complex DM candidate $\Phi^0_2$ and reduced it to one scalar and pseudoscalar, we can avoid such one loop scattering by choosing a mass splitting. In fact, one requires a non-zero mass splitting, at least greater than of the order of $\mathcal{O}(100 \; \text{keV})$, typical kinetic energy of DM particles, in order to avoid tree level inelastic scattering of DM off nuclei mediated by $Z$ boson \cite{inelasticIDM}.
\begin{figure}[htb]
\centering
\includegraphics[scale=0.41]{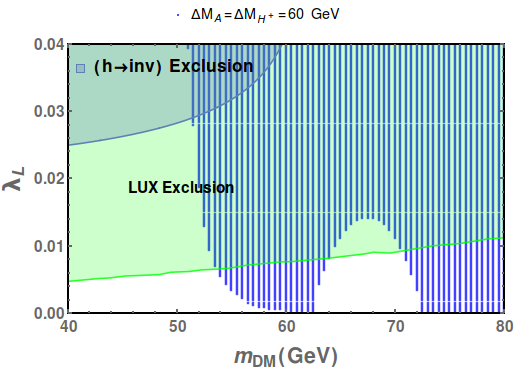}
\includegraphics[scale=0.45]{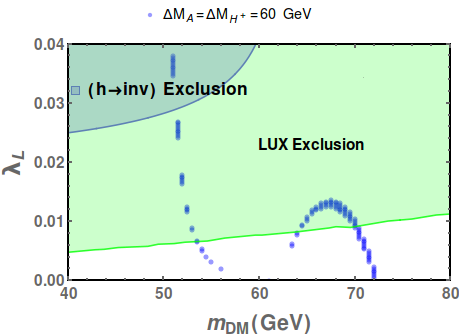}
\caption{Parameter space in the $\lambda_L-m_{DM}$ plane giving rise to dark matter relic abundance $\Omega_{\text{DM}} h^2 \leq 0.1187$ (left panel) and $ \Omega_{\text{DM}} h^2 \in 0.1187 \pm 0.0017$ (right panel) in pure IHDM.}
\label{fig2}
\end{figure}

We implement the model in micrOMEGA 4.3.1 \cite{micromega} to calculate the relic abundance of DM. We first reproduce the known results in IHDM by considering the neutral scalar $H$ to be the DM candidate having mass below the $W$ boson mass threshold. In the left panel of figure \ref{fig2}, we first show the parameter space of pure IHDM in $\lambda_L-m_{DM}$ plane that satisfies the condition $\Omega_{\text{DM}} h^2 \leq 0.1187$. We have taken both the mass difference $m_A-m_H=m_{H^{\pm}}-m_H=60 \; \text{GeV}$ as a typical benchmark value satisfying all other constraints. Such a large benchmark point reduces the coannihilation effects and show the dependence of relic abundance on Higgs portal coupling $\lambda_L$ in a visible manner.\footnote{We have not considered low mass differences in this work as that will make the coannihilations more efficient reducing the dependence of relic abundance on Higgs or lepton portal couplings and here our main motivation is to show the importance of lepton portal couplings.} The blue region in the left panel of \ref{fig2} therefore indicates the parameter space where the DM annihilation is either just enough or more than the required one to produce the correct relic abundance. Therefore, considering the additional lepton portal couplings for such values of $\lambda_L$ will further suppress the relic abundance. Therefore, we choose benchmark values of $\lambda_L-m_{DM}$ for our next analysis, from that region of this plot which overproduces the DM in pure IDM, so that an efficient lepton portal annihilation can bring down the relic abundance to the observed range. In the right panel of figure \ref{fig2}, we further impose the relic abundance criteria $ \Omega_{\text{DM}} h^2 \in 0.1187 \pm 0.0017$ which reduces the number of allowed points significantly from the one in the left panel. In both the plots we also show the LUX 2016 exclusion line based on the upper bound on DM nucleon scattering cross section.  We also show the LHC limit on Higgs invisible decay width which remains weaker than the LUX 2016 bound. The tiny allowed region near $m_{\text{DM}} \approx m_h/2$ corresponds to the s-channel resonance mediated by the SM Higgs while the allowed region of $m_{\text{DM}}$ close to $W$ boson mass threshold corresponds to the dominance of DM annihilation into three body SM final states mentioned above.

\begin{figure}
\centering
$
\begin{array}{cc}
\includegraphics[scale=0.45]{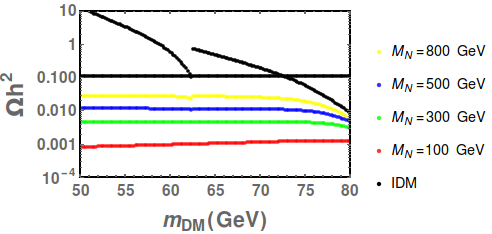} &
\includegraphics[scale=0.45]{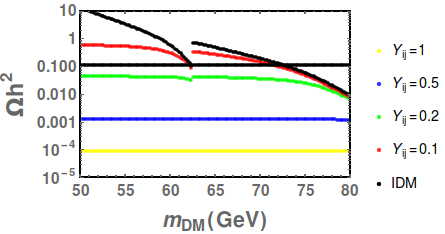} \\
\end{array}$
\caption{Effect of lepton portal couplings on dark matter relic abundance, for specific dark matter Higgs coupling $\lambda_L$. Left : Relic density vs. $m_{DM}$ for different $M_N$ with fixed Y=0.2. Right : Relic density vs. $m_{DM}$ for different Y with fixed $M_N$=1000 GeV.}
\label{fig3}
\end{figure}

%%%%%%%%%%%%%%%%%%%%%%%%%%%%%%
\begin{figure}
\centering
$
\begin{array}{cc}
\includegraphics[scale=0.45]{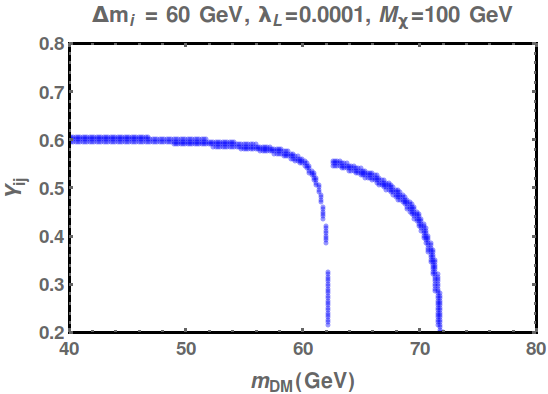}
\includegraphics[scale=0.46]{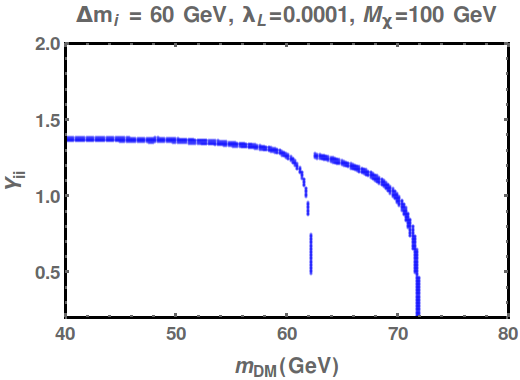}
\end{array}$
\caption{Parameter space in the $Y-m_{DM}$ plane giving rise to the correct dark matter relic abundance with 3$\sigma$ range for specific choice of $\lambda_L=0.0001$ and $M_{\chi}=100$ GeV. Left : nonzero off-diagonal Yukawa coupling scenario, Right : Diagonal Yukawa coupling scenario.}
\label{fig4}
\end{figure}
After reproducing the known results of IHDM in the low mass regime for a benchmark value of mass splitting, we calculate the DM relic abundance by incorporating the $Z_2$ odd heavy leptons. In figure \ref{fig3}, we show the effect of vector like neutral heavy leptons on relic abundance. To make DM annihilations through lepton portal more efficient, we choose the Higgs portal coupling to be very small $\lambda_L=0.0001$ and also keep both the mass splitting within the components of the inert scalar doublet as 60 GeV like before. In the left panel of figure \ref{fig3}, the effect of heavy neutral fermion mass on the relic abundance is shown for a fixed value of Yukawa coupling $Y=0.2$. In the right panel of figure \ref{fig3}, the effect of lepton portal Yukawa couplings on DM relic abundance is shown for fixed value of heavy neutral fermion mass $M_N=1000$ GeV. From both these panels of figure \ref{fig3}, it is clear that the leptonic portal can play a non-trivial role in generating the DM relic abundance. While the benchmark values of Higgs portal coupling and mass splitting chosen above produce correct DM abundance only for two different masses, the introduction of lepton portal can result in new allowed region of DM masses. As expected, the maximum effect of lepton portal on DM relic abundance occurs for smaller values of heavy lepton mass or equivalently large values of Yukawa couplings. Since neutral heavy fermion couplings with SM leptons are required to be tiny from neutrino mass constraints as can be seen from figure \ref{fig1a}, we consider only the effect of heavy charged leptons on DM relic abundance. The effect of charged lepton portal on DM relic abundance will be similar to that of neutral case discussed above.
\begin{figure}
\centering
$
\begin{array}{cc}
\includegraphics[scale=0.55]{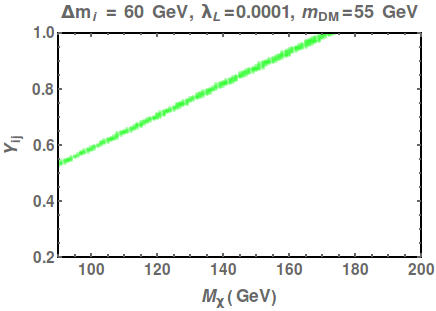} 
\includegraphics[scale=0.46]{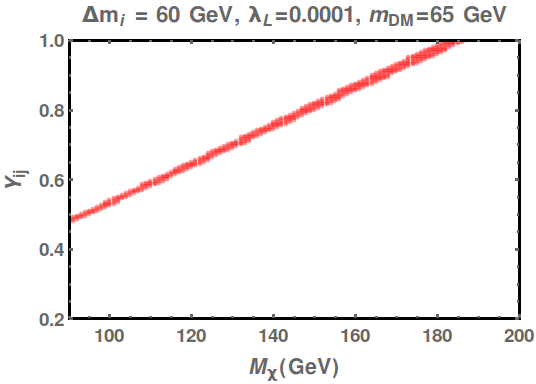}
\end{array}$
\caption{Parameter space in the $Y-M_{\chi}$ plane giving rise to the correct dark matter relic abundance with 3$\sigma$ range for specific choice of $\lambda_L=0.0001$ and $m_{DM}$ for nonzero off-diagonal Yukawa coupling scenario. Left : for $m_{DM}$=55 GeV, Right : for $m_{DM}$=65 GeV.}
\label{fig5}
\end{figure}

\begin{figure}
\centering
$
\begin{array}{cc}
\includegraphics[scale=0.53]{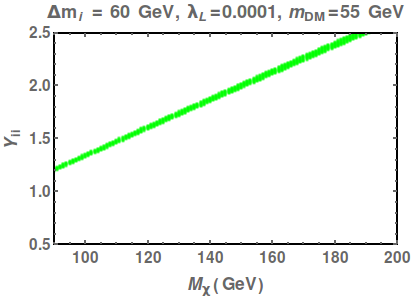} 
\includegraphics[scale=0.53]{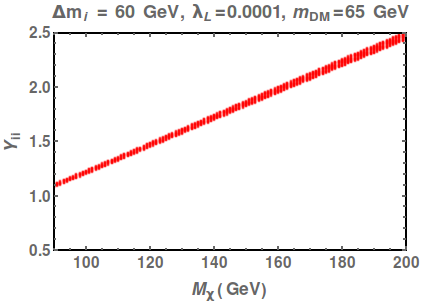}
\end{array}$
\caption{Parameter space in the $Y-M_{\chi}$ plane giving rise to the correct dark matter relic abundance with 3$\sigma$ range for specific choice of $\lambda_L=0.0001$ and $m_{DM}$ for Diagonal Yukawa coupling scenario. Left : for $m_{DM}$=55 GeV, Right : for $m_{DM}$=65 GeV.}
\label{fig6}
\end{figure}
After showing the effect of lepton portal on DM relic abundance for specific values of Yukawa and heavy neutral fermion masses, we do a general scan of these two parameters from the requirement of generating correct abundance. Since neutral heavy fermion portal is not efficient after neutrino mass constraints are incorporated, we do the general scan only for charged heavy lepton portal here. In figure \ref{fig4}, we show the allowed parameter space satisfying relic density in the $Y-m_{DM}$ plane for a benchmark point of IHDM parameters like before and taking the heavy charged fermion mass to be 100 GeV. The left panel of \ref{fig4} considers the lepton portal couplings to be of general non-diagonal type while the right panel considers the couplings to be diagonal. As discussed before, such diagonal couplings will evade the constraints from LFV decay. Since a diagonal structure of Yukawa couplings reduces the total number of annihilation channels, one requires larger values of Yukawa couplings to produce the correct relic abundance, compared to the ones in the non-diagonal case. In figure \ref{fig5}, we show the allowed parameter space in $Y-M_{\chi}$ plane for two specific dark matter masses $m_{DM}=55, 65$ GeV with general non-diagonal Yukawa couplings. The corresponding result for diagonal Yukawa couplings are shown in figure \ref{fig6}. It should be noted that these two benchmark values of DM masses in pure IHDM can not give rise to correct relic abundance for small values of Higgs portal couplings as seen from figure \ref{fig2}. However, after allowing the lepton portal couplings, we can generate correct relic abundance for such values of DM masses which remain disallowed in the pure IHDM.
\section{Collider Implications}
\label{sec4}
In pure IHDM, the pseudoscalar A can decay into $Z$ and $H$ whereas $H^{\pm}$ can decay to either $W^{\pm}  H$ or $W^{\pm} A$. When $m_{H^\pm}$ is close to $m_A$, then the first decay mode of $H^{\pm}$ almost dominates. Depending upon the decay mode of $W^{\pm}$ and $Z$, we have either pure leptonic plus missing transverse energy (MET) or hadronic plus MET or mixed final states from pair production of the inert scalars. Earlier studies in the IHDM \cite{Miao:2010rg,Gustafsson:2012aj,Datta:2016nfz} focussed on pair production of inert scalars and their decays into leptons and MET. In another recent work \cite{Poulose:2016lvz}, the authors studied dijet plus MET final states in the context of IHDM at LHC. The dilepton plus dijet plus MET and trilepton plus MET final states have also been studied in a recent work \cite{Hashemi:2016wup}. The 8 TeV constraints and 13 TeV projection from monojet plus MET are discussed in another work \cite{monojet_met}. 

In the presence of both $Z_2$ odd neutral and charged vector like leptons, additional channels open up. For example, now $H^{\pm}$ can decay to ${\chi}^{\pm}$ $\nu_{i}$ or $N_{i}$ $l^{\pm}$. Similarly, $A$ can decay into $l^{\pm}$ ${\chi}^{\mp}$ or $N_{i}$ $\bar \nu_i$. Since neutrino mass constraints push the mass of neutral leptons typically to the order of TeV range, both $H^{\pm}$ and A will mainly decay through charged vector like leptons (VLL)  ${\chi}^{\pm}$. Then  ${\chi}^{\pm}$ will further decay into $l^{\pm} H$. One can find earlier studies in the context of vector like leptons in references \cite{Nirakar_Paper,Nilanjana,VLL_limit,Soumya,Angelescu:2016mhl}. To highlight the difference in collider signatures with comparison to pure IHDM, we have considered a few benchmark points.
We choose the following benchmark points all of which correspond to the fixed values of $m_h$ = 125 GeV, $\lambda_L$ = 0.0001 , $\lambda_2$ = 0.1, $M_N$=1000 GeV, $Y$=0.001.
\begin{center}
BP1: $m_H$ = 55 GeV, $m_{H^+} = m_{A}$ = 115 GeV, $M_{\chi}$=100 GeV, $Y_{ii}$=1.5 \\
BP2: $m_H$ = 65 GeV, $m_{H^+} = m_{A}$ = 125 GeV,  $M_{\chi}$=100 GeV, $Y_{ii}$=1.5 \\
BP3: $m_H$ = 65 GeV, $m_{H^+} = m_{A}$ = 200 GeV,  $M_{\chi}$=150 GeV, $Y_{ii}$=2.0  \\
BP4: $m_H$ = 65 GeV, $m_{H^+} = m_{A}$ = 300 GeV,  $M_{\chi}$=150 GeV, $Y_{ii}$=2.0. 
\end{center}

In table \ref{table:cs_2lplusMET}, we have listed the parton level cross sections for final states that contribute to dilepton+MET final states at detector level in both IHDM and IHDM+VLL models for the above benchmark points. It should be noted that for BP1 and BP2, $H^\pm$ will go through off-shell decay that is, $H^\pm \to W^*{^\pm} H$ with $W^*{^\pm}$ decaying leptonically in pure IHDM case due to limited phase space availability.  But for BP3 and BP4, $H^\pm$ will go through on-shell decay that is, $H^\pm \to W{^\pm} H$ with $W{^\pm}$ decaying leptonically in pure IHDM case. In IHDM+VLL model, $H^\pm$ will decay to $\chi^\pm$ that is, $H^\pm \to{\chi}^{\pm}$ $\nu_l$ with $\chi^\pm$ further decaying into $l^{\pm} H$. It is clearly evident from this table that we have enhancement of the cross section in IHDM+VLL due to opening of new decay modes of $H^\pm$. We must highlight one point that it is very difficult to probe heavier charged Higgs mass (like the ones in BP3 and BP4) in pure IHDM case due to small cross section. But in the IHDM+VLL model discussed here, we have sufficient cross section to probe these heavier masses of charged Higgs. Apart from the channels listed in table \ref{table:cs_2lplusMET}, there is another process which contributes to dilepton plus MET final states that is $\chi^+\chi^-$ production with $\chi^{\pm}$ decays to ${l^\pm}H$. So as a whole, the dilepton plus MET final state will be an important collider signature to probe the modified IHDM that we discussed in this article. This inspires us to do a full signal versus background study at detector level which we will come up in a separate work \cite{2lmet}. 

%%%%%%%%%#######################Table-1#########################################################
\begin {table}[h!]
\begin{center}
\begingroup
\fontsize{10pt}{11pt}\selectfont
{
    \begin{tabular}{ | l | r | r| }
    \hline\hline
%\backslashbox{Types}{Fermions}
Benchmark&\multicolumn{2}{c|}{{$\sigma(p p \to H^{+} H^{-} \to 2l + 2\nu + 2H$)(in fb)}}\\ \cline{2-3}
Points &  IHDM  & IHDM+VLL   \\ \hline
BP1 & 8.1  & 126  \\ \hline
BP2   & 6.1 & 93.5 \\ \hline
BP3  & 1.7   &  13.8  \\ \hline
BP4  & 0.3   & 2.1 \\ \hline

\hline
  \end{tabular}
}
\endgroup
\caption{{ The parton level cross section for final states that contribute to dilepton+MET final states at detector level in both IHDM and IHDM+VLL models at the LHC ($\sqrt{s} = 14$ TeV) for different BPs considered. }}
\label{table:cs_2lplusMET}
\end{center}
\end{table}
\section{Conclusion}
\label{sec5}
We have studied a very specific region of parameter space in IHDM where the Higgs portal coupling of DM is very small, as suggested by null results in dark matter direct detection experiments so far. In the low mass regime of DM that is $m_{DM}<M_W$, such small value of Higgs portal coupling $\lambda_L$ may not be sufficient to produce the correct relic abundance of DM except for a a few specific values of $m_{DM}$. We then extend this model by heavy neutral and charged leptons which are also odd under the $Z_2$ symmetry of the IHDM. These heavy leptons can be motivating from neutrino mass as well as LHC phenomenology point of view, apart from their role in producing the correct DM relic abundance in those region of parameter space which can not produce correct relic in pure IHDM. The neutral heavy fermions can generate tiny neutrino masses at one loop level via scotogenic mechanism, requiring the corresponding Yukawa couplings to be small $(<10^{-4})$ for TeV scale heavy neutral fermion masses. This keeps the contribution of neutral heavy leptons to DM abundance suppressed. The heavy charged fermion couplings to DM are however, not constrained to be tiny from neutrino mass point of view and hence can be sizeable enough to play a role in DM abundance. We show that the entire low mass regime of IHDM is allowed from relic abundance criteria if the lepton portal parameters are suitably chosen. This does not affect the DM direct detection scattering rates as there are no tree level or one loop couplings of DM with nuclei through leptons. The heavy leptons can also give rise to observable LFV decay rates like $\mu \rightarrow e \gamma$ as well as interesting collider signatures like dilepton plus missing energy. Although for simplicity, we choose particular type of Yukawa structure which does not contribute to LFV decay rates, it is in principle possible to choose some structure of the Yukawa couplings which can simultaneously produce correct DM abundance as well as keep the decay rate of LFV decays like $\mu \rightarrow e \gamma$ within experimental reach. We also show how the lepton portal extension of IHDM enhances dilepton plus missing energy signals at the LHC, for chosen benchmark points. There can also be lepton number violating signal like same sign dilepton plus dijet plus missing energy in this model, but remain suppressed for the benchmark values chosen in our analysis.

\begin{acknowledgments}
We thank P. Poulose for useful discussions while carrying out this work. SS would like to thank Nirakar Sahoo for his constant help in resolving issues in micrOMEGA. Also SS thanks Biswajit Karmakar, Abhijit Saha for technical help in using mathematica and Ashis Kundu, Sourav Chattopadhyay for help in shell scripting. 

\end{acknowledgments}

\bibliographystyle{apsrev}

\end{document}